\documentclass[12pt,a4paper,preprintnumbers,superscriptaddress,showkeys]{revtex4-taml} 

\usepackage[fleqn]{amsmath}
\usepackage{amssymb}
\usepackage{bm}
\usepackage{graphicx} 
\usepackage{graphics}
\graphicspath{{Figures/}}
\usepackage{subfigure}
\usepackage[pdfstartview=FitH,
            CJKbookmarks=true,
            bookmarksnumbered=true,
            bookmarksopen=true,
            colorlinks=true,
            pdfborder=001,
            linkcolor=blue,
            citecolor=blue,
            urlcolor=blue
            ]{hyperref}
\RequirePackage{color}
\renewcommand{\thetable}{\arabic{table}} %
\renewcommand{\figurename}{Fig.} %
\renewcommand{\tablename}{Table}  %

\begin{document}

\title{Limit point buckling of a finite beam on a nonlinear foundation}

\author{R. Lagrange}
\affiliation{Massachusetts Institute of Technology, Department of Mathematics, Cambridge, MA 02139-4307, USA}

\date{\today}

\begin{abstract}
\noindent \textbf{Abstract}

{In this paper, we consider an imperfect finite beam lying on a
nonlinear foundation, whose dimensionless stiffness is reduced from
$1$ to $k$ as the beam deflection increases. Periodic equilibrium
solutions are found analytically and are in good agreement with a
numerical resolution, suggesting that localized buckling does not
appear for a finite beam. The equilibrium paths may exhibit a limit
point whose existence is related to the imperfection size and the
stiffness parameter $k$ through an explicit condition. The limit
point decreases with the imperfection size while it increases with
the stiffness parameter. We show that the decay/growth rate is
sensitive to the restoring force model. The analytical results on
the limit load may be of particular interest for engineers in
structural mechanics}.
\end{abstract}


\maketitle

\section{Introduction}

An elastic beam on a foundation is a model that can be found in a
broad range of applications: railway tracks, buried pipelines,
sandwich panels, coated solids in material, network beams, floating
structures... The usual way to model the interaction between the
beam and the foundation is to replace the latter with a set of
independent springs whose restoring force is a linear \cite[see
e.g.][]{Winkler1867,Lekkerkerker62,Naschie1974,Lee96,Kounadis2006,Koiter2009,Challamel2011,Suhir2012}
or a nonlinear \cite[see e.g.][]{Cox40,Potier15,Hunt93,
HuntBlack,Wadee97,Wadee18,Whiting17,Netto99,
Zhang2005,Jang2011,Lagrange2013} function of the local deflection of
the beam. In both cases, the nonlinear effects, from the beam's
deformation and/or from the restoring force, play a crucial role in
the buckling and the post-buckling behaviors. In particular, for a
softening nonlinear foundation, the equilibrium curves of the beam
may exhibit a maximum load (i.e. limit point) at which the structure
loses its stability. Small imperfections, arising from various
sources, usually have an appreciable effect on this maximum load.
The papers on deterministic imperfection sensitivity include those
of \cite[][]{Reissner70,Reissner71,Sheinman91,
Sheinman93,Hunt93,Whiting17,Lee96,Kounadis2006,Lagrange2012} and
extensive references for the stochastic imperfection sensitivity are
compiled in \cite[][]{Elishakoff2001}. As a general rule, the
maximum load at which the beam becomes unstable diminishes with
increasing imperfection size. Considering a finite beam on a
bi-linear/exponential foundation, \cite{Lagrange2012} has shown the
existence of a critical imperfection size ${A_{0c}}$ such that: if
$A_0<A_{0c}$, then the maximum load diminishes with the imperfection
size, from the critical buckling load predicted by the classical
linear analysis \cite[see][]{Potier15}, for $A_0=0$, to the Euler
load for $A_0=A_{0c}$. In this case, the decay rate is sensitive to
the restoring force model. For $A_0>A_{0c}$, the maximum load is the
Euler load (i.e. buckling load of a beam with no foundation).

In the present paper we aim to extend these results to two restoring
force models with more general softening behaviors. We derive an
analytical expression for $A_{0c}$ and study the evolution of the
maximum load with the imperfection size and the stiffness reduction.

\section{Formulation of the problem}

\begin{figure}
\begin{center}
\includegraphics[width=0.5\textwidth]{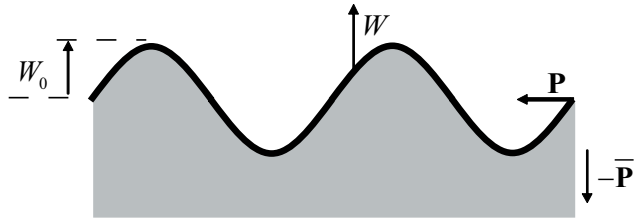}
\end{center}
\caption{Sketch of a beam on a nonlinear foundation. The beam has an
imperfect shape ${W_0}$ and its lateral displacement is $W$. The
compressive force is ${\bf{P}}$ and the restoring force per unit
length is ${-\bf{\overline{P}}}$.}\label{ProblemeModele}
\end{figure}

\begin{figure}
\begin{center}
\includegraphics[width=0.5\textwidth]{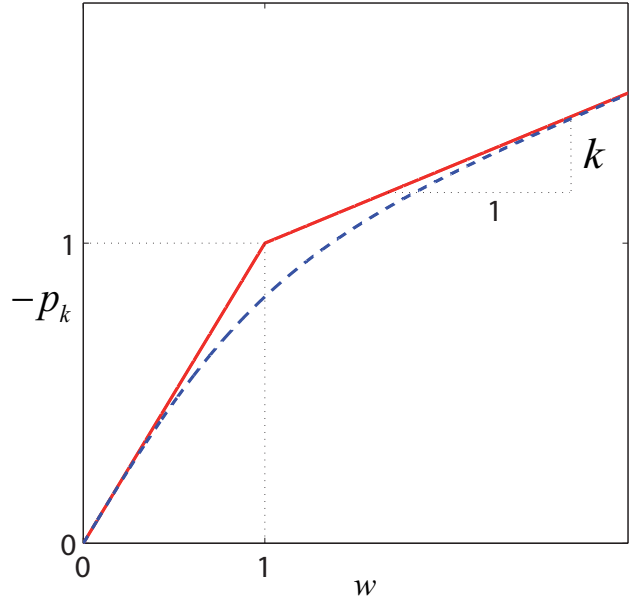}
\end{center}
\caption{Dimensionless restoring force $p_k$. Red line: bi-linear
model. Blue line: hyperbolic model. The stiffness ratio is
$k\leq1$.}\label{Fig2_4}
\end{figure}

We consider the effects of a compressive load $P$ on a beam of
length $L$, with bending stiffness $EI$, lying on a foundation that
provides a restoring force per unit length $\overline{P}$ (see Fig.
\ref{ProblemeModele}). The beam and the foundation are assumed to be
well bonded at their interface and remain bonded during deformation.
Thus, interfacial slip or debonding is not considered. The
mobilization of the foundation (also named the yield point) is noted
$\Delta$, its linear stiffness $K_0$ and its nonlinear stiffness
$K$.

In its initial configuration, the beam has an imperfect shape
$W_0=A_0\sin \left(\pi X/L\right)$, where $A_0$ is the imperfection
size and $X$ is the longitudinal coordinate.

We introduce the characteristic length $L_{c}  = \left(
{{{EI}}/{K_0}} \right)^{{1}/{4}}$ and the non-dimensional quantities
\begin{eqnarray}\label{GrandeursAdimBis2}
l &=& \frac{{ L }}{{L_{c} }},\quad x= \frac{{ X }}{{L_{c} }}, \quad
w = \frac{{W}}{{\Delta}}, \quad {w_0} = \frac{{{W_0}}}{{\Delta}},
\quad a_0=\frac{{A_0}}{\Delta}, \quad\lambda = \frac{P
{L_{c}}^2}{{EI }}, \quad k=\frac{K}{K_0}, \quad
p_k=\frac{\overline{P}}{K_0\Delta},\nonumber\\
\end{eqnarray}
as the dimensionless beam length, longitudinal coordinate, lateral
deflection (measured from the initial configuration), imperfection
shape, imperfection size, compressive load, stiffness ratio and
restoring force respectively.

Two models for the restoring force $p_k$ are considered in this
article (see Fig. \ref{Fig2_4}). The first one is
\begin{eqnarray}\label{RestoringForceDim}
p_k \left( w \right) =  - w - (1 - k)\left( {{\mathop{\rm sgn}}
\left( w \right) - w} \right){\rm{H}}\left( {\left| w \right| - 1}
\right),
\end{eqnarray}
where $\rm{sgn}$ denotes the sign function and $\rm{H}$ is the
Heaviside function, defined as ${\rm{H}}\left( \left| w \right| -
1\right)=0$ if $\left| w \right|<1$ and $1$ if $\left| w \right|>1$.
This bi-linear restoring force refers to a foundation whose
stiffness is instantaneously reduced from $1$ to $k\leq1$ when
$w>1$. The particular case $k=1$ corresponds to a linear foundation.
Reference \cite[][]{Lagrange2012} considered the particular case of
$k=0$. Here, we extend the study to $k\leq1$, which leads to more
general results.

To reflect the experimental tests on railway tracks performed by
\cite{Birmann}, also reported in \cite{Kerr2,Tvergaard14}, who
showed that the lateral friction force acting on a track is a smooth
function of the lateral displacement, we introduce a hyperbolic
profile defined as
\begin{eqnarray}\label{TanhDim}
p_k \left( w \right) =  - kw - (1 - k)\tanh (w).
\end{eqnarray}
This restoring force is a regularization of the bi-linear model as
they share the same asymptotic behaviors.

We assume that $\lambda$ and $p_k$ are conservative forces, that
strains are small compared to unity and that the kinematics of the
beam is given by the classical Euler-Bernoulli assumption. The
imperfection is also assumed to be small so that terms with higher
powers of $w_0$ or its derivatives are neglected in the expression
of the potential energy. Under these assumptions, the potential
energy $V$ with low-order geometrically nonlinear terms is
\cite[see][]{Potier15}

\begin{eqnarray}\label{EnergiePotentielle}
V = \int\limits_0^l {\left[ {\frac{1}{2}{w^{''}} ^2
-\lambda\,\left({\frac{1}{2}{w^{'}}^2 +{{w_0}^{'}}{{w}^{'}} }\right)
- \int\limits_0^w { p_k \left( t \right)dt} } \right]{{\text{d}x}}},
\end{eqnarray}
where a prime denotes differentiation with respect to $x$. The first
term in the integral is the elastic bending energy, the second is
the work done by the load $\lambda$, the last term is the energy
stored in the elastic foundation.

The equilibrium states are given by the critical values of $V$.
Assuming a simply supported beam, the boundary conditions are
$w\left( 0 \right)= w\left( l \right)=0$. Variations of
(\ref{EnergiePotentielle}) for an arbitrary kinematically admissible
virtual displacement $\delta w$ yields the weak formulation of the
equilibrium problem

\begin{eqnarray}\label{VariationEnergiePotentielle}
\int\limits_0^l \left[w^{''''}  + \lambda\,\left( {w^{''} + w_0
^{''} } \right) - {p_k} \left( w  \right) \right]\delta
w\,\text{d}x=0,
\end{eqnarray}
which is equivalent to the stationary Swift-Hohenberg equation

\begin{eqnarray}\label{EqDiff}
w^{''''}  + \lambda\,\left( {w^{''} + w_0 ^{''} } \right) - {p_k}
\left( w  \right)=0,
\end{eqnarray}
along with static boundary conditions $w''\left( 0 \right)=
w''\left( l \right)=0$.

This boundary value problem is nonlinear because of the restoring
force and its solutions are highly sensitive to the length $l$, as
shown in \cite[][]{Lee96}. Therefore, it is unrealistic to describe
the behavior of the system over a large range of variation for $l$.
As done in \cite{Lagrange2012}, this study is restricted to a finite
length beam where $l<\sqrt{2}\pi$. For such values of $l$ a
classical linear analysis \cite[see][]{Potier15} shows that the
first buckling mode is the most unstable one and appears for
$\lambda_c=\lambda_e+\lambda_e^{-1 }$, where $\lambda_e=\left( \pi
/l \right)^2$ is the Euler load.

\section{Solving methods}
To solve (\ref{VariationEnergiePotentielle}) we apply a Galerkin
method with a trigonometric test function $w$ of amplitude $y$: $w=y
\sin \left({\pi\,x }/{l}\right)$, assuming that the deflection has
the same shape as the first buckling mode and the initial
imperfection. For more details about the principle of the method,
the reader is referred to \cite{Lagrange2012}, where the procedure
has already been used. In that paper, this method has been shown to
be reliable in the prediction of the equilibrium paths of the
system, for $k=0$. We shall see in the present paper that it is
actually reliable for any $k\leq1$, thereby extending the results of
\cite{Lagrange2012}.

The insertion of $\delta w=\delta y \sin \left({\pi\,x }/{l}\right)$
in (\ref{VariationEnergiePotentielle}) yields

\begin{eqnarray}\label{IntegraleWhiting}
\int\limits_0^l {\sin \left( {\frac{\pi }{l}x} \right)\left[
{w^{''''}  + \lambda \left( {w^{''}  + w_0 ^{''} } \right) -
p_k\left( w \right)} \right]{{\text{d}x}} = 0}.
\end{eqnarray}
Splitting the restoring force in a linear and a nonlinear term
$N\left(w\right)$ leads to
$p_k\left(w\right)=-w-\left(1-k\right)N\left(w\right)$. With this
decomposition and  $w=y \sin \left({\pi\,x }/{l}\right)$,
(\ref{IntegraleWhiting}) can be rewritten as

\begin{eqnarray}\label{Galerkin}
\lambda_k =\frac{1}{a_{0} +y} \left[\lambda_c\,
y+\left(1-k\right)\frac{Q\left(y\right)}{\lambda_{e} } \right],
\end{eqnarray}
where the subscript $k$ denotes the dependance of $\lambda$ on the
parameter $k$. The function $Q$ takes into account the nonlinear
behavior of the restoring force and is given by

\begin{eqnarray}\label{1.7}
Q\left(y\right)=\frac{2}{l} \int_{0}^{l}\sin \left(\frac{\pi }{l}
x\right)N\left(y\sin \left(\frac{\pi }{l}
x\right)\right){{\text{d}x}}.
\end{eqnarray}

For the restoring force models (\ref{RestoringForceDim}) and
(\ref{TanhDim}), $Q$ is negative and decreases monotonically to the
asymptote $-y+{4}/{\pi}$ as $y \to +\infty$. Thus $\lambda_k$ is
maximum for $k=1$ (linear foundation) and has an horizontal
asymptote $\lambda _k^\infty= \lambda _e + k\lambda _e ^{ - 1}$ as
$y \to +\infty$.

Equilibrium paths predicted by (\ref{Galerkin}) are traced out in
the plane $\left(y={\rm{max}}\left(w\right), \lambda\right)$ by
gradually incrementing $y$ and evaluating $\lambda_k$, $k$ and $a_0$
being fixed. Predictions are compared with a numerical solution of
(\ref{EqDiff}), using MATLAB's boundary value solver {\it{bvp4c}}
\cite[this is a finite difference code that implements a collocation
formula, details of which can be found in][]{Matlab}.

\section{Results}
\begin{figure}
\begin{center}
\includegraphics[width=1.\textwidth]{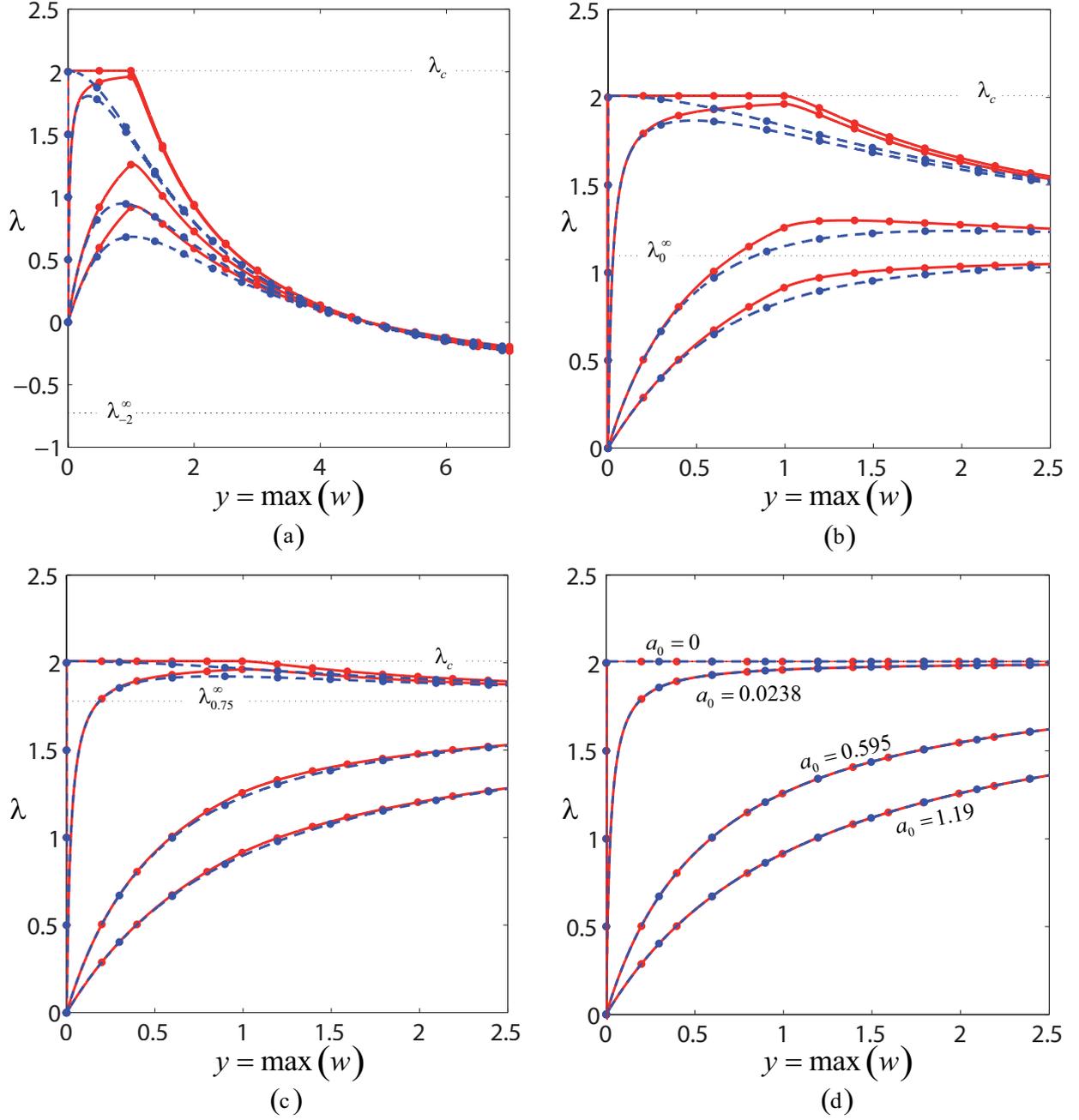}
\end{center}
\caption{Equilibrium paths of a finite length beam on a nonlinear
foundation. Circles: numerical predictions. Lines: Galerkin
solution. In red: bi-linear restoring force model. In blue:
hyperbolic model. (a) $k=-2$, (b) $k=0$, (c) $k=0.75$, (d) $k=1$. On
each subfigure, the equilibrium paths are plotted (from top to
bottom) for $a_0=0$, $a_0=0.0238$, $a_0=0.595$ and $a_0=1.19$, as
shown in (d). The length of the beam is $l=3$. }\label{Paths}
\end{figure}

\begin{figure}
\begin{center}
\includegraphics[width=0.5\textwidth]{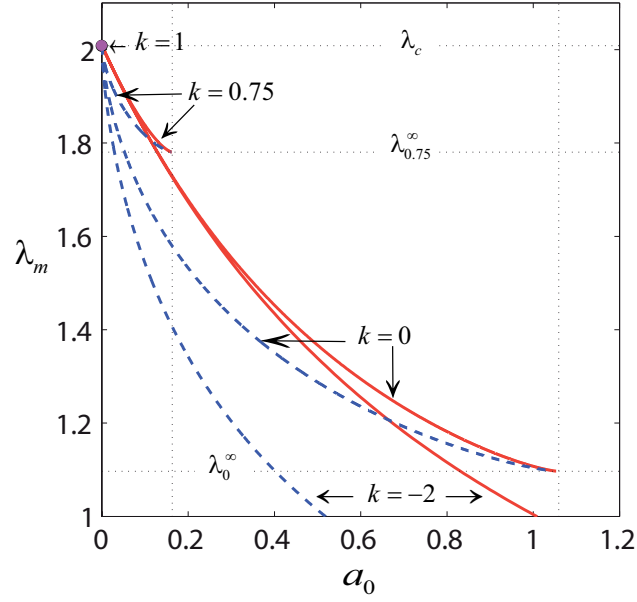}
\end{center}
\caption{Maximum load $\lambda_m$ that the beam may support versus
the imperfection size $a_0$ and the foundation stiffness ratio $k$.
In red: bi-linear restoring force model. In blue: hyperbolic model.
Vertical dotted lines correspond to the critical imperfection sizes
$a_{0c}$ for $k=0.75$ and $k=0$. The length of the beam is $l=3$.
}\label{Limit_point}
\end{figure}

\begin{figure}
\begin{center}
\includegraphics[width=0.5\textwidth]{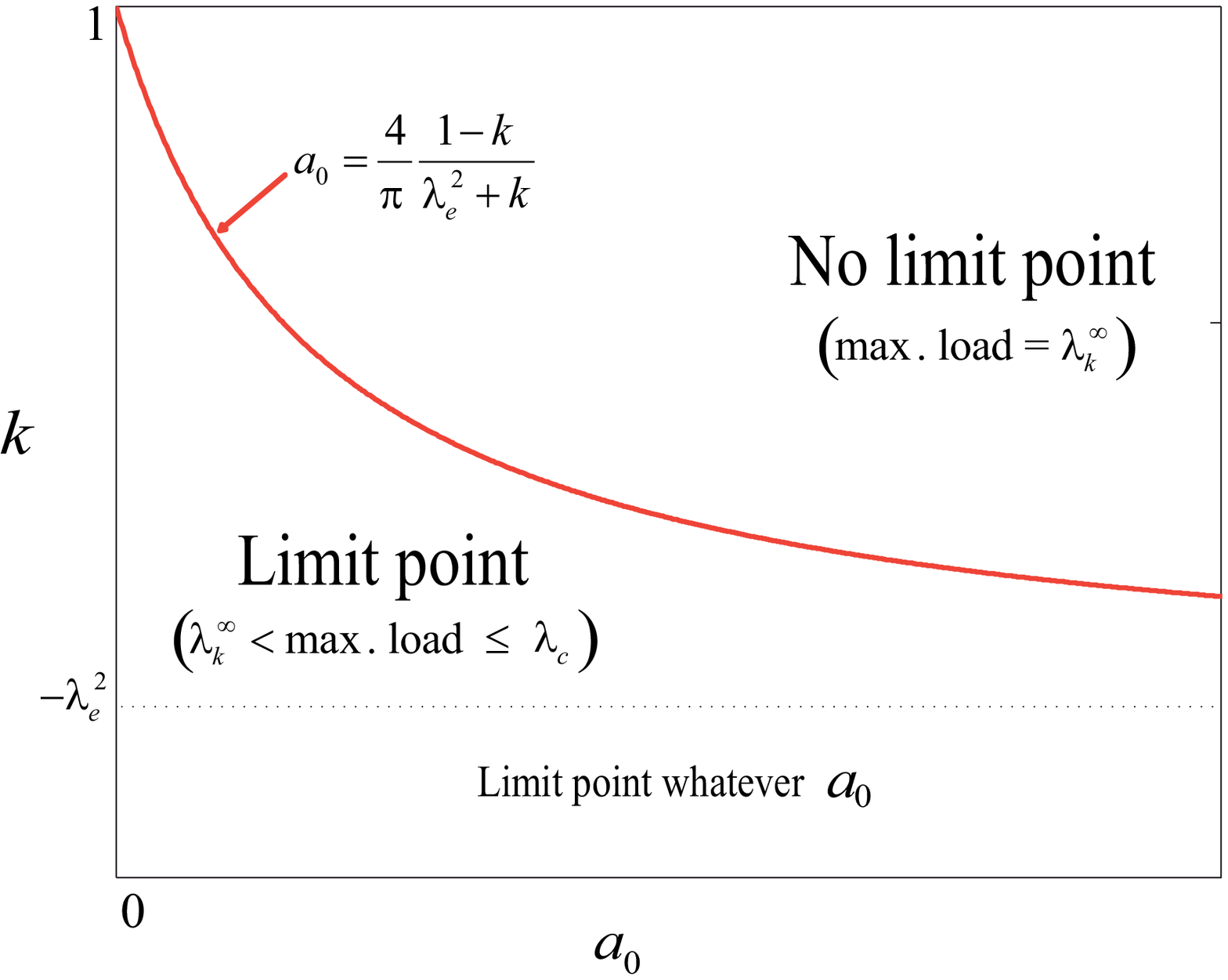}
\end{center}
\caption{Diagram of existence of a limit point for an imperfect
finite beam on a bi-linear/hyperbolic foundation. $k$ is the
stiffness ratio of the foundation and $a_0$ the imperfection size.
$\lambda_e=(\pi/l)^2$ is the Euler load,
$\lambda_c=\lambda_e+\lambda_e^{-1}$ and $\lambda_{k}^{\infty}=
\lambda_e+k\lambda_e^{-1}$.}\label{Diagram}
\end{figure}

The equilibrium paths predicted by the Galerkin method and the
numerical solution are shown in Fig. \ref{Paths}. A perfect
agreement in the predictions is found for both restoring force
models (the relative error between the two methods being less than
$0.1\%$). Since the Galerkin method was initiated with a test
function having the same shape as the imperfection, we conclude that
the deflection is just an amplification of the initial curvature. In
other words, in the range $l<\sqrt{2}\pi$, no localized buckling is
observed for a beam on a bi-linear or hyperbolic foundation. This
behavior has also been reported by \cite{Lee96} for a linear
foundation, showing a tendency toward localization when increasing
the beam length.

As expected, the equilibrium paths traced out for the hyperbolic
restoring force are below those traced out for the bi-linear force,
the hyperbolic profile modeling a softer foundation than the
bi-linear one. However, the choice of the restoring force has little
influence on the shape of the equilibrium paths.

For small $a_0$, the equilibrium paths first increase to a maximum
$\lambda_m$ that is smaller (or equals to in the case of $a_0=0$)
than the buckling load $\lambda_c$. Then, the paths asymptotically
decrease to $\lambda_{k}^{\infty}$. In the case of $a_0=0$, $k=1$,
they remain equal to $\lambda_c$. For high $a_0$, the equilibrium
paths increase monotonically to the asymptote
$\lambda_{k}^{\infty}\leq\lambda_c$. The asymptotic value
$\lambda_{k}^{\infty}=\lambda_c$ is reached for $a_0=0$ and $k=1$.

Note that for $k<-{\lambda_e}^2$, $\lambda_{k}^{\infty}$ is
negative, so that equilibrium states with $\lambda<0$ are predicted
(see Fig. \ref{Paths}(a)). Physically, when $k<-{\lambda_e}^2$ the
restoring force $p_k$ may become negative so that springs are
compressed, pushing up the beam. In this situation, the restoring
force has a destabilizing effect on the beam. To counteract this
effect, a tensile force $\lambda<0$ has to be applied.

The evolution of $\lambda_m$ versus $a_0$ is shown in Fig.
\ref{Limit_point}. A gradual drop in the maximum load admissible by
the structure from $\lambda_c$ to $\lambda_{k}^{\infty}$ is observed
when increasing $a_0$ (resp. decreasing $k$). This gradual drop is
highly sensitive to the restoring force model. A log scale applied
on Fig. \ref{Limit_point} shows that, for small imperfection sizes,
the decay rate does not depend on $k$: $\lambda _m - \lambda _c$
scales as $-a_0$, for the bi-linear model and as $-a_0 ^{2/3}$ for
the hyperbolic model.

For $a_0$ larger than a critical value $a_{0c}$, the equilibrium
paths do not have a limit point anymore. Actually, a path with no
limit point may be seen as a path with a limit point at
$\left(\infty,\lambda_{k}^{\infty}\right)$. Thereby, $a_{0c}$ may be
obtained from (\ref{Galerkin}) by enforcing $y\rightarrow\infty$ in
$d\lambda_k/dy=0$. Both restoring force models leads to

\begin{eqnarray}\label{a0c}
a_{0c}=\frac{4}{\pi}\frac{1-k}{{\lambda_e}^{2}+k},
\end{eqnarray}
whose dimensional equivalent form is

\begin{eqnarray}\label{A0_c}
A_{0c}  = \frac{4}{\pi }\frac{{\left( {K_0  - K} \right)\Delta
}}{{{{\pi ^4 EI}}/{{L^4 }} + K}}.
\end{eqnarray}

The critical imperfection size predicted by \cite{Lagrange2012} is
therefore recovered in the particular case $K=0$, showing that
$A_{0c}$ only depends on the limiting plateau $K_0\,\Delta$ of the
restoring force \cite[as stated in][]{Maltby7}.

Finally, since $a_0>0$, equation (\ref{a0c}) shows that if
$k<-{\lambda_e}^2$ then the equilibrium paths always have a limit
point $\lambda_{k}^{\infty}<\lambda_m<\lambda_c$.

\section{Conclusion}

In this paper, we considered the buckling of an imperfect finite
beam on a bi-linear/hyperbolic foundation. The imperfection has been
introduced as an initial curvature of size $a_0$ and the foundation
stiffness ratio as a parameter $k\leq1$, extending the result of
\cite{Lagrange2012} derived for $k=0$.

Equilibrium paths of the beam have been predicted using a Galerkin
method initiated with a single trigonometric function which has the
same shape as the imperfection. Predictions compare well with a
numerical solution and lead to the conclusion that only periodic
buckling can arise for a finite beam on a bi-linear/hyperbolic
foundation, as also observed by \cite{Lee96} for an linear
foundation.

We have shown the existence of a critical imperfection size $a_{0c}
= 4\left( {1 - k} \right)\left[ {\pi \left( {\lambda _e^2  + k}
\right)} \right]^{ - 1}$, independent of the restoring force model,
such that:

\begin{itemize}
    \item if $a_0<a_{0c}$, then the maximum load diminishes with
    increasing imperfection size, from $\lambda _c=\lambda_e+\lambda_e^{-1}$, for
$a_0=0$, to $\lambda_e+k\lambda_e^{-1}$ for $a_0=a_{0c}$,
$\lambda_e=(\pi/l)^2$ being the Euler load. The decay rate has been
shown to be sensitive to the restoring force model. In the limit of
small $a_0$, $\lambda _m - \lambda _c\sim -a_0$ for the bi-linear
model and $\lambda _m - \lambda _c\sim -a_0 ^{2/3}$ for the
hyperbolic model.
    \item If $a_0>a_{0c}$, then the
maximum load simply corresponds to $\lambda_e+k\lambda_e^{-1}$.
\end{itemize}

Finally, we have shown that if $k<-{\lambda_e}^2$ then an imperfect
finite beam on a bi-linear/hyperbolic foundation can support a
compressive load larger than $\lambda_{k}^{\infty}$, and smaller
than $\lambda_c$, whatever the imperfection size is. This feature is
highly interesting for an engineer since $a_0$ is usually hard to
evaluate. The main results from this study are summarized in Fig.
\ref{Diagram}.

In the present paper, a bi-linear restoring force model for the
foundation has been used but plasticity effects that would emerge
from loading/unloading cycles have not been considered. Future works
will have to highlight the way those effects could modify the
maximum load that the beam can support. A basic model would consist
of considering a permanent deflection as an imperfection whose size
would grow up after each cycle. In that case, from the present
study, it is expected a decrease of the maximum load after each
cycle, at least as long as the accumulated deflection remains
smaller than a threshold equivalent to $a_{0c}$.

The author acknowledges Dr. M. Brojan for introducing to him the
hyperbolic restoring force model and Dr. Alban Sauret and Dr. Jay
Miller for their insightful comments on this paper.
\bibliographystyle{aipnum4-1}
\bibliography{Biblio}


\end{document}